# Optically Controlled Jitter Generator

Julia Manasson and V. A. Manasson

(WaveBand Corporation, Torrance, California)

*Abstract--* A new simple circuit producing random pulse trains is proposed and experimentally studied. The circuit is composed of an operational amplifier and two feedback links, one of which comprises two photodiodes. The photodiodes are responsible for nonlinearity in the feedback. By varying the illumination it is possible to control the nonlinearity in the photodiode current-voltage characteristics and change the degree of randomness in the oscillations. The circuit's simplicity and optical control make it attractive for coupled map lattices.

*Index terms:* nonlinear circuits, chaos generator, optical control.

## I. INTRODUCTION

Networks comprised of nonlinear circuits produce spatio-temporal patterns and constitute an attractive prototyping medium for studying complex systems [1]. Of special interest are networks based on chaotic oscillators. To build large networks it is usually necessary for network components, their control, and their read-out to be simple and impermeable to parasitic cross-talk. This is particularly true of chaotic oscillators, which are super-sensitive to interference. In this paper we propose a new chaotic circuit. It is simple, it produces robust jitter, and most importantly, it can be controlled by optical means. Networks based on the proposed generator can either be globally controlled (by using a single optical source) or individually controlled (by using optical fibers) independent of how the cells are coupled. In both cases parasitic interference from controlling circuitry is minimal.

## II. CIRCUIT IMPLEMENTATION

A chaotic circuit should comprise at least one nonlinear component. In our circuit, a tandem of two photodiodes connected in series toward one another plays this role. The current-voltage characteristic of a typical photodiode is given by [3]:

$$I = I_0[\exp(qV/kT) - 1] - q\Phi \qquad (1),$$

where $k$ is Boltzman's constant, $T$ is the absolute temperature, $q$ is the electron charge, $I$ is the current through the photodiode, $V$ is the voltage across the device, $\Phi$ is the photonic flux, and $I_0$ is the dark saturation current. A diode tandem I-V characteristic is provided by the modification of equation (1) and given by

$$I = (q\Phi + I_0)\tanh(qV/2kT) \qquad (2).$$

It represents a nonlinear function that is symmetrical with respect to the origin (see Fig. 1) and saturates at high voltages.

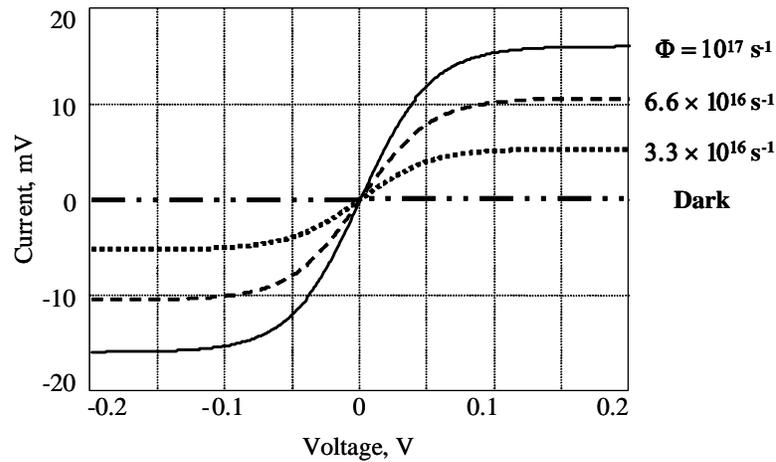

**Fig. 1**

The saturation current depends on the illumination level. The maximum conductance occurs at the origin and is likewise proportional to the illumination level.

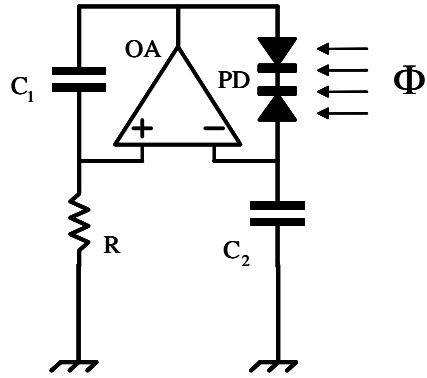

**Fig. 2**

The circuit is shown in Fig. 2. It represents a modified relaxation generator. Adding a nonlinear link comprised of a photodiode tandem dramatically changes the nature of oscillations. In contrast to a conventional relaxation generator that produces periodical pulse trains, the circuit produces robust jitter. We observed random pulse trains with a wide variety of circuit components, including capacitors and resistors of different values, various types of operational amplifiers, and encapsulated Schottky diodes in place of photodiodes. Chaotic behavior was absent only when the photodiodes were replaced with linear resistors.

## III. MEASURED RESULTS

We studied how illumination affects the circuit behavior. Experimental data presented in this paper was obtained with the following circuit components: resistor $R = 550\Omega$, capacitors $C_1 = 0.12$ μF and $C_2 = 0.12$ μF, operational amplifier (OA) TL084OCA, and photodiodes (PD) TOX9100 (Texas Optoelectronics). In the experiments the photodiodes were illuminated by daylight. Different illumination levels were obtained by covering the circuit with light absorbing materials. To estimate the flux $\Phi$ that was incident on the photosensitive areas of the diodes, we measured the photocurrent (which is proportional to $\Phi$) generated by a photodiode identical to and placed near the photodiodes in tandem. The measurements were obtained at eight different illumination levels: $\Phi_1$, $\Phi_2=3.1\Phi_1$, $\Phi_3=8.7\Phi_1$, $\Phi_4=23\Phi_1$, $\Phi_5=50\Phi_1$, $\Phi_6=110\Phi_1$, $\Phi_7=260\Phi_1$, $\Phi_8=1100\Phi_1$, and in the dark.

Fig. 3 shows typical waveforms produced by the circuit in two extreme cases: in the dark (Fig. 3a) and at the highest illumination level (Fig 3b). In the dark the waveform represents a sequence of pulses with approximately the same pulse duration and a random temporal position. At the highest illumination the waveform sequence is almost repeatable (coherent). Variations in the pulse train "period" recorded for illumination levels between the two extreme cases are shown in Fig. 4. Each pulse train comprises 17 consecutive pulses. The strongest period variation occurs in the dark, and deviation from the average period reaches 50% or more. Randomness gradually disappears with increasing illumination.

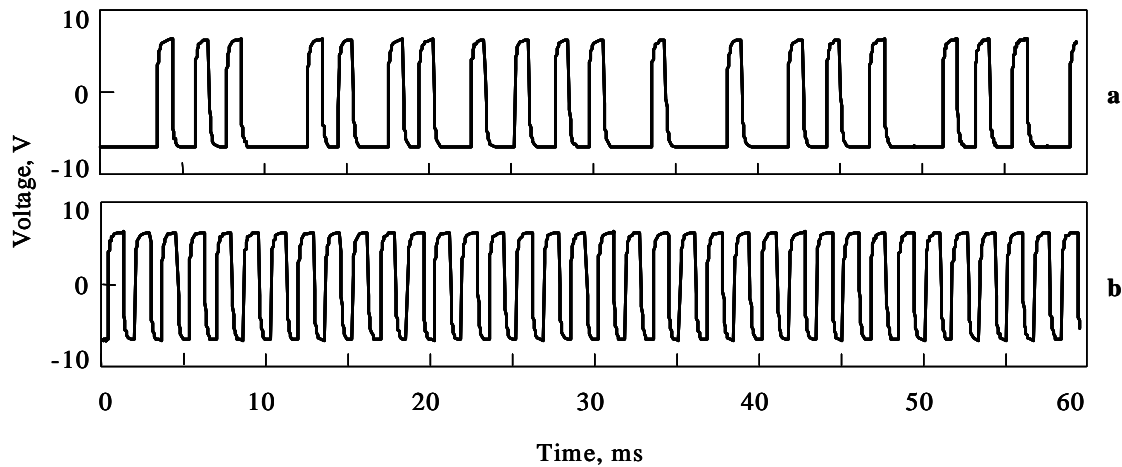

**Fig. 3**

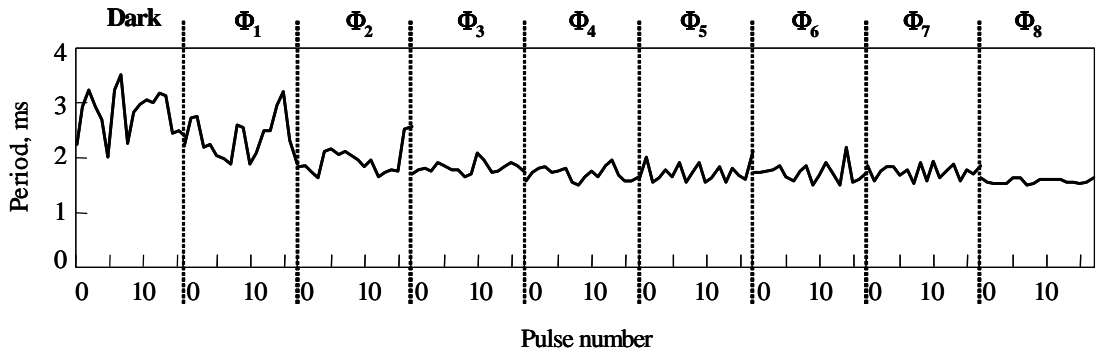

Fig. 4

Using experimental data we calculated auto-correlation functions that play the role of Lyapunov exponents in characterizing the degree of randomness. Three auto-correlation functions corresponding to different illumination levels are shown in Fig. 5.

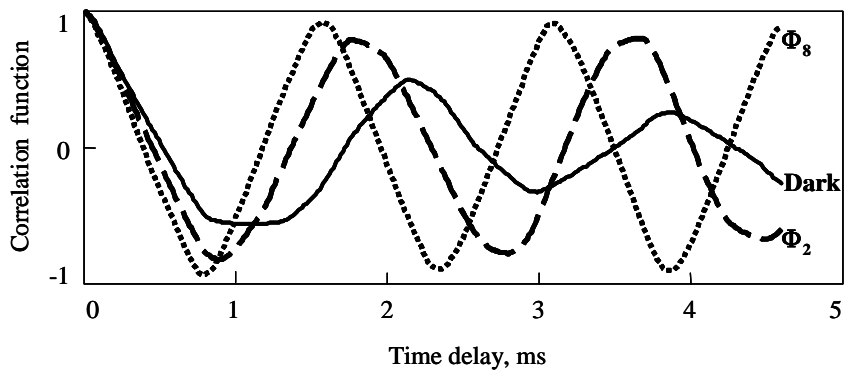

Fig. 5

In all the curves, correlation between same-sequence pulses decays with time, but the rate of decay varies for different illumination levels. The higher the illumination, the slower is the divergence. The power spectrums confirm this behavior. Three power spectrums are shown in Fig. 6.

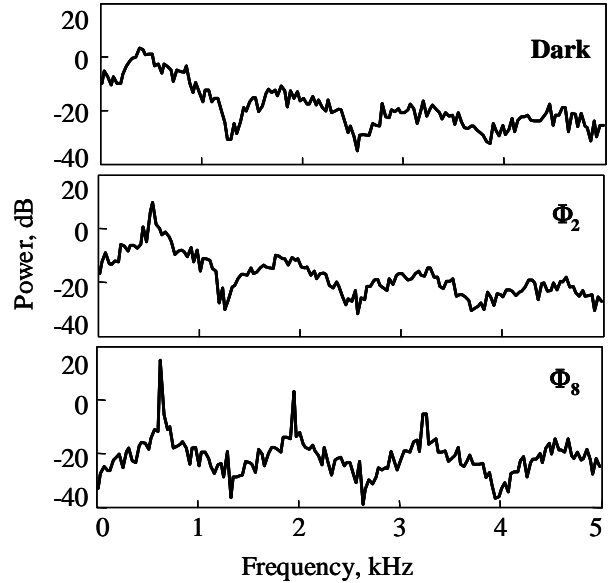

**Fig. 6**

At the highest illumination level there are several discrete peaks and relatively low background noise. As illumination decreases, the peaks become wider and spread out across the spectrum. In the dark (the upper curve) the spectrum is almost continuous, which is typical of chaotic oscillations.

IV. CONCLUSIONS

We have proposed a new simple circuit that produces randomly repeated pulse trains. The degree of randomness can be regulated by circuit illumination. This is demonstrated by the waveforms, power spectrums, and auto-correlation functions. Due to its simplicity and ability to produce robust chaos, the proposed jitter generator may be useful for a number of chaos-based applications. Optical control creates a basis for new architectures of networks. Examples include networks with well-developed (branched) optical links and networks with global optical control, both of which are well protected from electromagnetic interference.